\documentclass[conference]{IEEEtran}
\IEEEoverridecommandlockouts
\usepackage{cite}
\usepackage{amsmath,amssymb,amsfonts}
\usepackage{algorithmic}
\usepackage{graphicx}
\usepackage{textcomp}
\usepackage{xcolor}
\def\BibTeX{{\rm B\kern-.05em{\sc i\kern-.025em b}\kern-.08em
    T\kern-.1667em\lower.7ex\hbox{E}\kern-.125emX}}
    
\usepackage{graphics} 
\usepackage{epsfig} 
\usepackage{amsmath} 
\usepackage{amssymb}  
\usepackage{color}  

\usepackage{cite}

\newcommand {\qed}{\hfill\mbox{\rule[0pt]{1.3ex}{1.3ex}}}

\newtheorem{assum}{Assumption}
\newtheorem{lem}{Lemma}
\newtheorem{rem}{Remark}

\begin{document}

\title{Accelerated Consensus  for Multi-Agent Networks \\ through Delayed Self Reinforcement 
\thanks{Funding from NSF grant CMMI 1536306 is gratefully acknowledged}}

\author{\IEEEauthorblockN{ Santosh Devasia}
\IEEEauthorblockA{\textit{Department of Mechanical Engineering} \\
\textit{University of Washington}\\
Seattle, USA \\
devasia@uw.edu}
}

\maketitle
\thispagestyle{empty}
\pagestyle{empty}

\begin{abstract}
This article aims to improve the performance 
of networked multi-agent systems, which are common representations of 
cyber-physical systems. 
The rate of convergence to consensus  of multi-agent networks is critical to 
ensure cohesive, rapid response to external stimuli. 
The challenge is that increasing the rate of convergence 
can  require changes in the network connectivity, 
which might not be always feasible. 
Note that current consensus-seeking control laws can be considered as a gradient-based search over the graph's Laplacian potential. 
The main contribution 
of this article is to improve the convergence to consensus, 
by using an accelerated gradient-based search approach. 
Additionally, this work shows that  the 
accelerated-consensus approach can be implemented in a distributed 
manner, where each agent applies a delayed self reinforcement, 
without the need for additional network information or changes to the 
network connectivity. 
Simulation result shows that the convergence 
rate  with the accelerated consensus 
is about double the convergence rate of  current consensus laws. 
Moreover, the loss of synchronization during the transition 
is reduced by about ten times with the use of the proposed 
accelerated-consensus approach. 
\end{abstract}

\vspace{0.1in}

\section{Introduction}
\vspace{-0.01in}
Multi-agent networks are common cyber-physical systems with applications such as
autonomous vehicles, swarms of robots and other unmanned systems, e.g., ~\cite{Huth_92,Vicsek_95,Jadbabaie_03,Ren_Beard_05,Olfati_Saber_06,Karafyllis_15,Peng_IJC_16,Meng_IJC_17}. 
The performance of such systems, such as the response to external stimuli, depends on 
rapidly transitioning from one operating point (consensus value) to another, e.g., as seen in flocking,~\cite{Attanasi_14,Wang_IJC_15}. Thus, there is interest to increase the convergence 
to consensus for such networked multi-agent systems. 

\vspace{0.1in}
A challenge is that there are fundamental limits to the achievable convergence to consensus using existing graph-based update laws 
for a given network, e.g., of the form 
\begin{align}
Z[k+1]   =   \left[  {\textbf{I}} -\gamma K \right] Z(k)  = P Z[k]
\label{EQ_intro_standard}
\end{align}
where the 
current state is $Z(k)$, the updated state is $Z[k+1]$,  $\gamma$ is the update gain, $K$ is the graph Laplacian and $P$ is the Perron matrix. Hence the 
convergence to consensus depends on the  
eigenvalues of the Perron matrix $P$, which in turn depends on the 
eigenvalues of the graph Laplacian $K$. 
For example, if the underlying graph  is undirected and connected, it is well known that convergence to consensus can be achieved provided the update gain $\gamma$ is  
sufficiently small, e.g.,~\cite{Olfati_Murray_07}. 
The gain $\gamma$ can be selected to maximize the convergence rate. 
However, for a given graph  (i.e., a given graph Laplacian $K$), the range of the acceptable update gain $\gamma$ 
is limited, which in turn, limits the achievable rate of convergence as shown in previous work~\cite{Devasia_2019_ICC}. 
Although  it is possible to change the convergence by choosing the Perron matrix~\cite{Chen_Weisheng_14}, i.e., by choosing a different graph structure for the network, the maximum rate of convergence with current 
graph-based updates is bounded for a given network structure. 

\vspace{0.1in}
As discussed in~\cite{Devasia_2019_ICC}, current limitations in graph-based approaches motivate the development of  new approaches to improve the convergence to consensus. Note that the convergence can  be slow if the number of agent inter-connections is small compared to the number of agents, e.g.,~\cite{Carli_08}. 
Randomized time-varying connections can lead to faster convergence, as shown in,   e.g.,~\cite{Carli_08}. The update sequence  of the agents can also be arranged to improve convergence, e.g.,~\cite{Fanti_15}.
When such time-variations in the graph structure or selection of the graph Laplacian $K$ are not feasible, 
the need to maintain stability 
limits the range of acceptable update gain $\gamma$, and therefore, 
limits the rate of convergence. 
This convergence-rate limitation motivates the proposed effort to develop a new approach to improve the network performance.

\vspace{0.1in}
The major contribution of this work is to use an accelerated-gradient-based approach to modify the standard
update law in Eq.~\eqref{EQ_intro_standard} for networked multi-agent systems. Previous works have used such acceleration methods (also referred to as the Nesterov's gradient method) to improve the convergence of gradient-based search in learning algorithms, e.g., see 
~\cite{Rumelhart_86,QIAN1999145}.
Another contribution is to show that the proposed 
accelerated approach to consensus  can be implemented by 
using a delayed self reinforcement (DSR), 
where each agent only uses 
current and past information from the network. 
This use of already existing information is advantageous since the 
consensus improvement is achieved without the need to change the network connectivity 
and without the need for additional information from the network. 
This work generalizes the author's previous works in~\cite{Devasia_2019_ICC,devasia_19_arXiv,Devasia_2018_DSCC}, which considered a 
momentum term only to improve the convergence to consensus. 

\newpage
\vspace{0.1in}
Simulation results of an example networked system are presented in this work to show  
that the proposed accelerated-consensus approach with DSR can substantially improve synchronization during the transition by about ten times, in addition 
to decreasing the transition time by about half, 
when compared to the case without the DSR approach.
This is shown to improve formation control during transitions in networked 
multi-agent systems.

\vspace{0.1in}
\section{Problem formulation}
\vspace{-0.01in}

\subsection{Background: network-based consensus control}
\vspace{-0.01in}

Let the multi-agent network be modeled using a graph representation, where 
the connectivity of the agents is represented by 
a directed graph (digraph) 
${\cal{G}} = \left({\cal{V}}, {\cal{E}}\right)$, e.g., as defined in~\cite{Olfati_Murray_07}. 
Here, the  agents are represented by nodes $ {\cal{V}}= \left\{ 1, 2, \ldots, {n\!+\!1} \right\}$, $n>1$ and their connectivity by edges $ {\cal{E}}   \subseteq {\cal{V}} \times {\cal{V}} $, where each agent $j$ belonging to the set of 
neighbors $N_i  \subseteq {\cal{V}} $  of the agent $i$ satisfies  $ j \ne i$ and $(j,i) \in {\cal{E}} $.

\vspace{0.1in}
\subsection{Graph-based control}
The consensus control for the multi-agent network is defined 
by the graph ${\cal{G}}$, as 
\begin{align}
\hat{Z}[k+1] & =  \hat{Z}[k] +\gamma  u(k) \nonumber \\
&  = \hat{Z}[k] +  \gamma  \sum_{i,j =1}^{n}  a_{ij}\left( \hat{Z}_j[k] - \hat{Z}_i[k] \right) \nonumber \\
& = \hat{Z}[k] -  \gamma  L \hat{Z}[k]  \nonumber \\
& = \left( {\bf{I}} - \gamma  L  \right) \hat{Z}[k] 
\label{system_eq}
\end{align}
where $\hat{Z}$ represents the states of the agents, 
$k$ represents the time instants $t_k = k \delta_t$, 
 $\gamma $ is the update gain, 
$u$ is the input to each agent, the weight $a_{i,j}$ is nonzero (and positive)
if and only if $j$ is in the set of neighbors $N_i  \subseteq {\cal{V}} $  of the agent $i$, 
and the terms
 $l_{ij}$ of the $(n+1)\times(n+1)$ Laplacian $L$  of the graph ${\cal{G}}$ are real and given by 
\begin{eqnarray}
\label{eq_laplacian_defn}
l_{ij} & =   \left\{ 
\begin{array}{ll}
-a_{ij} < 0, 	& {\mbox{if}} ~ j \in N_i \\
\sum_{m=1}^{n+1} a_{im}, & {\mbox{if}} ~ j = i,  \\
0 & {\mbox{otherwise,}}
\end{array}  
\right.
\end{eqnarray}
where 
each row of the Laplacian $L$ adds to zero, i.e., from Eq.~\eqref{eq_laplacian_defn}, 
the $(n+1) \times 1$ vector of ones  ${\textbf{1}}_{n+1}= [1, \ldots, 1]^T$  is a right eigenvector of the Laplacian $L$ with eigenvalue $0$,   
\begin{eqnarray}
\label{eq_first_lambda_eigenvector}
L{\textbf{1}}_{n+1} & = 0  {\textbf{1}}_{n+1} .
\end{eqnarray}

\subsection{Network dynamics}
One of the agents is assumed to be a virtual source agent, which can be 
used to specify a desired consensus value $Z_d$.  Without loss of 
generality, the last node, $n+1$ is assumed to be a virtual source agent. Moreover, each 
agent in the network should have access to the source agent's state ${Z}_s = \hat{Z}_{n+1}$  through the network, as formalized below. 

\vspace{0.1in}
\begin{assum}[Connected graph]
\label{assum_digraph_properties}
The digraph ${\cal{G}}$ is assumed  to have a directed path from the source node ${n+1}$ to any other  node $i$ in the graph, i.e.,  $ i \in {\cal{V}} \setminus \!{(n+1)}$. 
\qed 
\end{assum}

\vspace{0.1in}
Some properties of the  graph  ${\cal{G}}$ without the source node $n+1$ 
are listed below, e.g.,~\cite{Olfati_Murray_07}. 
In particular, consider the $n \times n$ pinned Laplacian  matrix $K$ 
obtained by removing the row and column associated with the source node $n+1$,  
the following partitioning of the Laplacian $L$ is invertible, i.e., 
\begin{eqnarray}
\label{eq_K_eigenvector}
L & = 
\left[
\begin{array}{c|c}
K  &  -B  \\
 \hline
  \star_{1 \times n} & \star_{1 \times 1} 
\end{array}
\right] ~\quad {\mbox{with}} ~~
\det{(K)} \ne 0, 
\end{eqnarray}
and $B$ is an  $n \times 1$ matrix 
\begin{equation}
\label{eq_B_def}
\begin{array}{rl}
B & = [ a_{1,s}, a_{2,s}, \ldots, a_{n,s}]^T
 \\ &  
 ~= [ B_{1}, B_{2}, \ldots, B_{n}]^T , 
\end{array}
\end{equation}
Non-zero values of $B_j$ implies that the agent $j$ is 
directly connected to the source $Z_s$.

\begin{enumerate}
\item 
The  pinned Laplacian matrix $K$  is invertible from the Assumption~\ref{assum_digraph_properties} and the 
 Matrix-Tree Theorem in~\cite{tuttle_graph}. 
 \item 
 The eigenvalues of $K$ have have strictly-positive, real parts.
 \item 
 The product of the inverse of the pinned Laplacian $K$ with $B$ leads to a  $n \times 1$ vector of  ones, i.e., 
\begin{eqnarray}
\label{eq_KinvtimesB}
K ^{-1} B & = {\textbf{1}}_n.
\end{eqnarray}
\end{enumerate}

\vspace{0.1in}
The dynamics of the non-source agents 
$Z$ represented by the remaining graph ${\cal{G}}\!\setminus\!s$,  be  given by 
\begin{equation}
\begin{array}{rl}
Z[k+1] &  = Z[k]  -\gamma  K Z[k] + \gamma   B  Z_s[k]
 \\
&  = \left(   {\bf{I}} - \gamma  K     \right) Z[k]   + \gamma   B  Z_s (k) \\
& = P  Z[k] + \gamma   B  Z_s (k) .
\label{system_non_source}
\end{array}
\end{equation}
where $P$ is Perron matrix.

\vspace{0.1in}
A sufficiently small selection of the update gain $\gamma $ will stabilize the dynamics in Eq.~\eqref{system_non_source}, e.g., see~\cite{Olfati_Murray_07}, i.e., 
all eigenvalues $\lambda_{P,i}$ of the Perron matrix 
$$P   ~ =  {\textbf{I}}_{n\times n}-\gamma  K,$$
where $ {\textbf{1}}_{n\times n}$ is the $n\times n$ identity matrix, 
will lie inside the unit circle. 

\subsection{Stable consensus}
With a stabilizing  update gain $\gamma$, the state $Z$ of the network (of all non-source agents)  converges to a fixed source value $Z_s$, e.g., 
for a step change in the source value $Z_s$,  i.e., $Z_s[k] = Z_d$ for $k > 0$ and zero otherwise. 
Since the eigenvalues of $P$ are inside the unit circle, the solution to Eq.~\eqref{system_non_source} for the step input 
converges 
\begin{eqnarray}
\left[ Z[k+1] - Z[k] \right] &  = P^{k}  \left[ Z[1] -  Z[0] \right]  \rightarrow 0 
\label{Eq_controlled_gen_soln}
\end{eqnarray}
as $ k \rightarrow \infty$.
Therefore, taking the limit as $k \rightarrow \infty$  in Eq.~\eqref{system_non_source}, and from invertibility of the pinned Laplacian $K $ from 
Eq.~\eqref{eq_K_eigenvector}. 
\begin{eqnarray}
Z[k]  \rightarrow K^{-1} B  Z_d
\label{Eq_controlled_gen_soln_2}
\end{eqnarray}
 as $ k \rightarrow \infty$.  
Then, from Eq.~\eqref{eq_KinvtimesB}, 
the state $Z[k]$  at the non-source agents reaches 
the desired state $Z_d$ as time step $k$ increases, i.e, 
\begin{eqnarray}
Z[k]  &    \rightarrow {\textbf{1}}_n Z_d ~~{\mbox{as}}~~  k \rightarrow \infty.
\label{system_non_source_stability}
\end{eqnarray}
Thus, the control law in Eq.~\eqref{system_non_source} achieves consensus. 

\vspace{0.1in}
\subsection{Convergence-rate limit} 
\label{Limit_of_convergence_rate}
For a given pinned Laplacian $K$, the range of the acceptable update gain $\gamma$ 
is limited, which in turn limits the achievable rate of convergence. 
If $$\lambda_{K,m}=m_{K,m} e^{i  \phi_{K,m}} $$ 
is an eigenvalue of the pinned Laplacian $K$ with 
a corresponding eigenvector $V_{K,m}$, i.e., 
\begin{eqnarray}
K  V_{K,m}   ~ & =    \lambda_{K,m}V_{K,m}, 
\label{eq_stability_condition_lem_Stability_and_Update_gain_pr_1}
\end{eqnarray}
then $$ \lambda_{P,m} = 1- \gamma \lambda_{K,m }$$ is an eigenvalue of the Perron matrix $P$ for the same eigenvector  $V_{K,m}$, since 
\begin{eqnarray}
P V_{K,m}  ~ =  \left[ {\textbf{I}}_{n\times n}-\gamma K \right]  V_{K,m} 
= (1 - \gamma \lambda_{K,m}) V_{K,m}.
\label{eq_stability_condition_lem_Stability_and_Update_gain_pr_2}
\end{eqnarray}

\vspace{0.01in}
\begin{lem}[Perron matrix properties]~ \hfill \\
\label{lem_Stability_and_Update_gain}
The network dynamics in Eq.~\eqref{system_non_source}, is stable if and only if 
the update gain $\gamma$ satisfies  
\begin{eqnarray}
0 ~ < \gamma ~ & <  \min_{1\le i \le n}  2   \frac{ \cos{(\phi_{K,i})} }{m_{K,i}}  = \overline\gamma  ~ < ~ \infty .
\label{eq_stability_condition_lem_Stability_and_Update_gain}
\end{eqnarray}
\end{lem}

\noindent
{{Proof:}} See~\cite{Devasia_2019_ICC}.
\hfill \qed 

\vspace{0.1in}
The model in Eq.~\eqref{system_non_source}  can be rewritten as 
\begin{equation}
\begin{array}{rl}
\frac{Z[k+1]- Z[k]} {\delta_t} &  =   -\frac{\gamma}{\delta_t}  K Z[k] + \frac{\gamma}{\delta_t}  B  Z_s [k]
\end{array}
\end{equation}
where $\delta_t$ is the  time between updates. 
For a sufficiently-small update time $\delta_t$ it can 
considered as the discrete version of the continuous-time  dynamics 
\begin{equation}
\begin{array}{rl}
\dot{Z}(t) &  =   -\frac{\gamma}{\delta_t} K Z (t) + \frac{\gamma}{\delta_t}   B  Z_s (t).
\label{system_non_source_contnuous}
\end{array}
\end{equation}
The eigenvalues of $ \frac{\gamma}{\delta_t} K $  increase proportionally with $\gamma$ and inversely with update time interval $\delta_t$. Therefore, the settling time $T_s$ of the continuous time system decreases as the gain $\frac{\gamma}{\delta_t}$ increases. 
The sampling time $\delta_t$ is bounded from below based on the
 sensing-computing-actuation bandwidth of the agents in the network, and 
 the gain $\gamma$ is limited by the network structure as in Lemma~\ref{lem_Stability_and_Update_gain}. 
Consequently, the  smallest possible update time $\delta_t$ and the given network structure 
limit the fastest possible settling time for a given network.

\subsection{The settling-time improvement problem}
\vspace{-0.01in}
The research problem addressed in this article is to   reduce the settling time $T_s$   (from one consensus state to another) 
under step changes in the source value (i.e., improve convergence) where each agent can modify  its update law 
\begin{enumerate}
\item 
using only existing information from the network neighbors, 
\item 
without changing the network structure (network connectivity $K$), and  
\item 
without changing the update-time interval $\delta_t$, which limits the maximum gain $\gamma$.
\end{enumerate}

\vspace{0.1in}
\section{Proposed accelerated consensus aproach}

\vspace{0.1in}
\subsection{Graph's Laplacian potential}
For undirected graphs, the control law $u$ in  Eq.~\eqref{system_eq} can be considered as a  gradient-based search based on the 
graph's Laplacian potential $\Phi_{{\cal{G}}}$~\cite{OlfatiSaberIEEETAC_04}, 
\begin{align}
u(\hat{Z}) & = -{\frac{1}{2}}\nabla \Phi_{{\cal{G}}}(\hat{Z}), 
\label{Eq_u_gradient}
\end{align}
where~\cite{OlfatiSaberIEEETAC_04,Zhang_hui_IJC_2015}
\begin{align}
\Phi_{{\cal{G}}}(\hat{Z}) & =   \frac{1}{2} \sum_{i,j =1}^{n} a_{ij}\left( \hat{Z}_j - \hat{Z}_i \right)^2
\label{system_eq_potential}
\end{align}
results in 
\begin{align}
u(\hat{Z}) & = -{\frac{1}{2}}\nabla \Phi_{{\cal{G}}}(\hat{Z}) ~ = -L \hat{Z}. 
\label{Eq_u_gradient_2}
\end{align}
This results in,  from   Eq.~\eqref{system_eq}, 
\begin{align}
\hat{Z}[k+1] & =   \hat{Z}[k]   -  \gamma {\frac{1}{2}} \nabla \Phi_{{\cal{G}}}(\hat{Z}) \nonumber \\
& =   \hat{Z}[k] -  \gamma  L \hat{Z}[k] 
\label{system_gradient}
\end{align}

\vspace{0.1in}
\subsection{Accelerated gradient search}
In general, the convergence of the gradient-based approach as in Eq.~\eqref{Eq_u_gradient}
can be improved using accelerated methods. In particular, applying  
the Nesterov modification~\cite{Rumelhart_86,QIAN1999145} of the traditional gradient-based method to  
 Eq.~\eqref{Eq_u_gradient} results in 
\begin{align}
u(\hat{Z}[k]) & = -{\frac{1}{2}}\nabla \Phi_{{\cal{G}}}\left\{\hat{Z}[k] +\beta \left(\hat{Z}[k] -\hat{Z}[k-1]  \right)  \right\} \nonumber \\\ 
& \qquad  + \beta \left(\hat{Z}[k] -\hat{Z}[k-1]  \right)   \nonumber \\ 
& = -L \left\{\hat{Z}[k] +\beta \left(\hat{Z}[k] -\hat{Z}[k-1]  \right)  \right\}  
\nonumber \\\ 
& \qquad  + \beta \left(\hat{Z}[k] -\hat{Z}[k-1]  \right) .
\label{Eq_u_accelerated_gradient}
\end{align}
This accelerated-gradient-based input
results in the modification of the system Eq.~\eqref{system_eq}  to 
\begin{align}
\hat{Z}[k+1] & =  \hat{Z}[k]   -  \gamma  L\left( \hat{Z}[k] + \beta\left(\hat{Z}[k]-\hat{Z}[k-1] \right) \right)   
\nonumber \\
& \qquad
+ \gamma  \beta\left( \hat{Z}[k]-\hat{Z}[k-1] \right) .
\label{System_Nesterov_gradient_approach}
\end{align}

\vspace{0.1in}
\noindent
Consequently,  the dynamics of the non-source agents 
$Z$ represented by the remaining graph ${\cal{G}}\!\setminus\!s$ and 
given by Eq.~\eqref{system_non_source}, becomes
\begin{equation}
\begin{array}{rl}
Z[k+1] &  = Z[k]   -\gamma  K \left( Z[k]  + \beta\left({Z}[k]-{Z}[k-1] \right) \right)   \\
&    \qquad + \beta\left({Z}[k]-{Z}[k-1] \right) + \gamma   B  Z_s[k]  .
\end{array}
\label{accelerated_system_non_source}
\end{equation}

\vspace{0.1in}
\begin{rem}
For directed graphs, the potential in Eq.~\eqref{system_eq_potential} 
does not lead to the graph Laplacian~\cite{OlfatiSaberIEEETAC_04,Zhang_hui_IJC_2015}. 
Instead, the graph potential (without the source node) can 
be directly considered as 
\begin{align}
\Phi_{{\cal{G}}\!\setminus\!s}({Z}) &  = {Z}^T K {Z} + B  Z_s, 
\label{system_eq_potential_digraph}
\end{align}
and the application of the  accelerated-gradient approach
leads to the same Eq.~\eqref{accelerated_system_non_source}. 
\hfill \qed
\end{rem}

\vspace{0.1in}
\subsection{Implementation using delayed self reinforcement}
\begin{figure}[!ht]
\begin{center}
\includegraphics[width=.95\columnwidth]{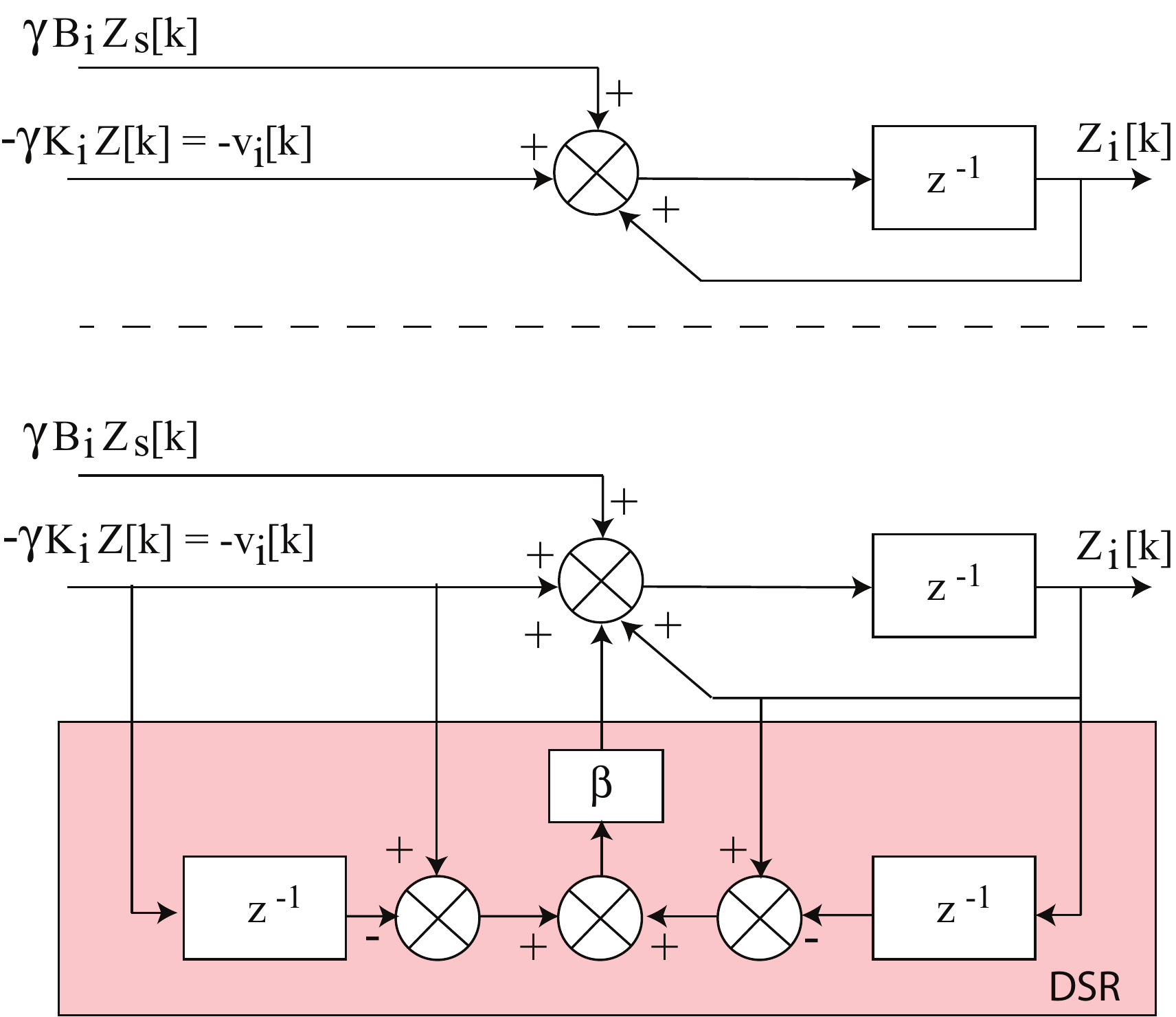}
\vspace{0.01in}
\caption{(Top) Implementation of standard gradient-based approach to multi-agent networks for the  $i^{th}$  agent as in Eq.~\eqref{system_non_source}. 
(Bottom) Delayed self reinforcement (DSR) approach to implement the accelerated-gradient-based approach for the $i^{th}$ agent  in Eq.~\eqref{accelerated_system_non_source}
without using additional network information.
}
\label{fig_1_control_implementation}
\end{center}
\end{figure}

The above accelerated-gradient approach for multi-agent networks 
can be implemented without additional information from the network, 
or having to change the network connectivity. 
For an agent $i$, let $v_i$ be the information obtained from the network, i.e., 
\begin{equation}
\begin{array}{rl}
v_i[k]  &  =  \gamma K_i  Z[k]  \\
\end{array}
\label{accelerated_system_non_source_2}
\end{equation}
where $K_i$ is  the $i^{th}$ row of the pinned Laplacian $K$.
Then, the update  of agent $Z_i$ is,  from Eq.~\eqref{System_Nesterov_gradient_approach}, 
\begin{equation}
\begin{array}{rl}
Z_i[k+1] &  = Z_i[k]   -\gamma  K_i \left( Z[k]  + \beta\left({Z}[k]-{Z}[k-1] \right) \right)   \\
&    \qquad + \beta\left({Z}_i[k]-{Z}_i[k-1] \right) + \gamma   B_i  Z_s[k]   \\[0.05in]
& = Z_i[k] - \left(v_i[k]  +\beta(v_i[k] -v_i[k-1]) \right) \\
&  \qquad + \beta\left({Z}_i[k]-{Z}_i[k-1] \right) +  \gamma B_i Z_s[k] 
\end{array}
\label{accelerated_system_non_source}
\end{equation}
where  $B_i$ is the $i^{th}$ row of the source connectivity matrix 
$B$. 
The delayed self-reinforcement (DSR) approach, however, requires each agent to store a delayed versions $Z_i[k-1]$ and $ v_i[k-1]$ of its current  state $Z_i[k]$ and information $v_i[k]$ from the network, 
as illustrated in Fig.~\ref{fig_1_control_implementation}.  ~ \hfill \qed

\suppressfloats

\subsection{Quantifying synchronization during transition}
In general, it is not only important that the network reaches  a new consensus value $Z_d$, but also that during the transition 
the network states are similar to each other. 
For example, consider the case when 
the state $Z$ of the agents are horizontal velocities $V_x$. Then having similar velocities 
during the transition (i.e., synchronization during the transition) can aid in maintaining the formation, without the 
need for additional control actions. 

\vspace{0.1in}
The lack of  cohesion or synchronization during the transition can be quantified in terms of the 
deviation $\Delta$  in the response 
as 
\begin{align}
\Delta &  =  \frac{\delta_t}{Z_d} \sum_{k=1}^{k_{Ts} } \left| Z[k] - \overline{Z}[k]  \right|_1
\label{Eq_synchronization}
\end{align}
where $k_{Ts} $ is the number of steps needed to reach the settling time $T_s$, which 
is time by which all agent responses $Z$ reach and stay within $2\%$ of the final value $Z_d$,  
$ \overline{Z}$ is the average value of the state $Z$, over all individual agent state-components $z_i$, i.e., 
\begin{align}
 \overline{Z}[k]  & = \frac{1}{n} \sum_{i=1}^{n} Z_i[k], 
\label{Eq_synchronization_2}
\end{align}
and 
 $|  \cdot |_1  $ is the standard vector 1-norm, 
$$| \hat{Z} |_1  = \sum_{i=1}^{n} | \hat{Z}_i|$$ for any vector $\hat{Z}$. 
A normalized measure ${\Delta}^*$  that removes the effect of the response speed is obtained by 
dividing the expression in Eq.~\eqref{Eq_synchronization} with the settling time $T_s$ as 
\begin{align}
{\Delta}^* &  =   \frac{\Delta}{T_s} .
\label{Eq_norm_synchronization}
\end{align}
Note that the system's transient response is more synchronized if the normalized deviation ${\Delta}^* $ is small.

\section{Results and discussion}
\label{Results_and_Discussion}
\vspace{-0.01in}
The step response of an example system, with and without DSR, are comparatively evaluated. Moreover, the impact of using DSR  on the response of a networked formation of agents is illustrated when the networked state is the velocity during an acceleration maneuver. 

\subsection{System description}
\begin{figure}[!ht]
\begin{center}
\includegraphics[width=.75\columnwidth]{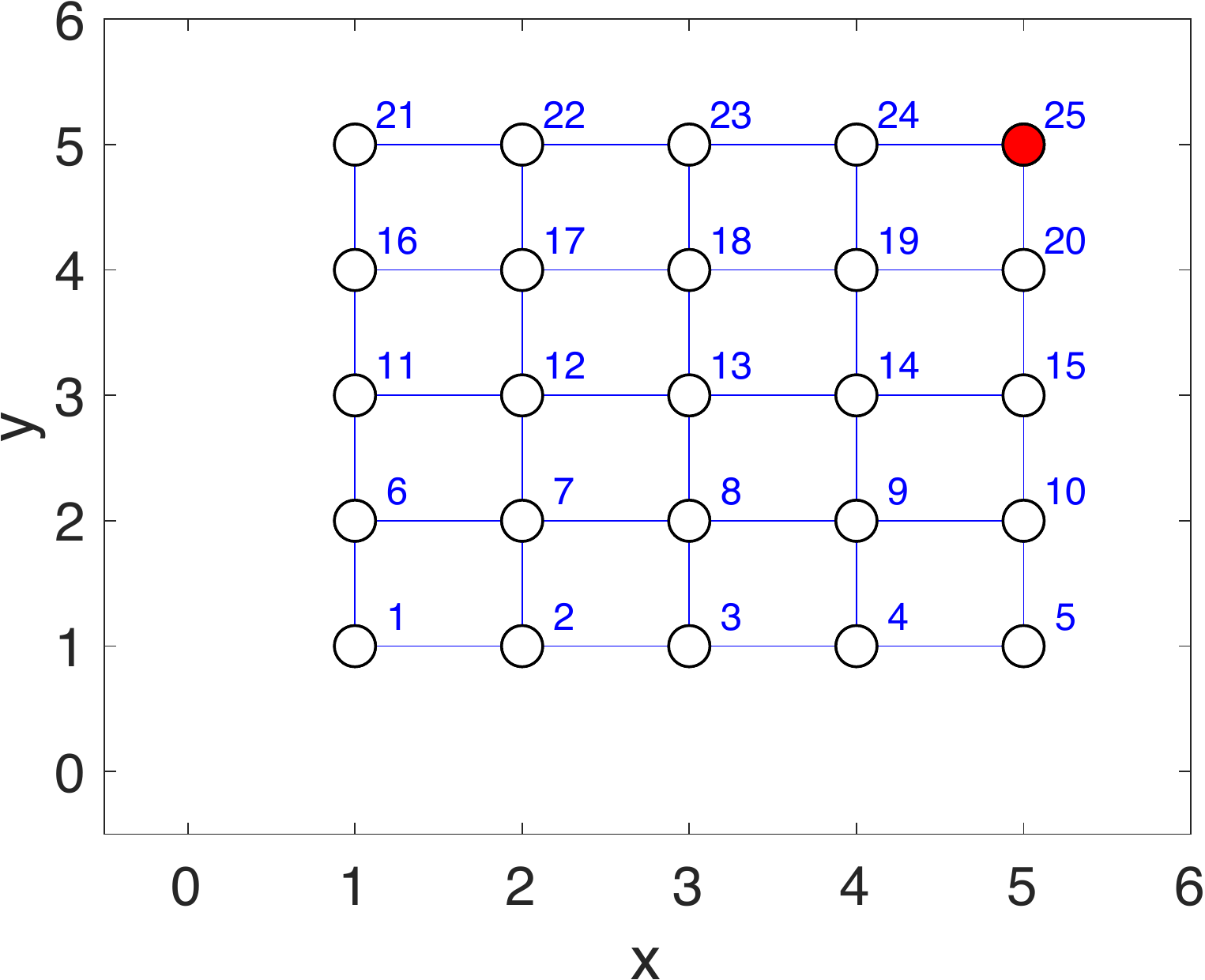}
\vspace{0.01in}
\caption{Graph ${\cal{G}}\!\setminus\!s$ of the non-source agent network used in simulations. The leader, shown in red, has 
access to  the source $Z_s$. 
}
\label{fig_example_graph}
\end{center}
\end{figure}

The example network used in the simulation is shown in Fig.~\ref{fig_example_graph}. It consists of $n=25$ non-source agents arranged  uniformly (initially)
on a $5\times 5$ grid. The minimal initial spacing between the agents is one. 
The last non-source agent $Z_{n}$ has access to the source $Z_s$. 

\suppressfloats

\vspace{0.1in}
The update gain $\gamma$ of the system in Eq.~\eqref{system_non_source} 
without DSR is selected to ensure stability. 
The weight of each edge is selected as one, i.e., $a_{ij}=1$ in Eq.~\eqref{eq_laplacian_defn}. 
The maximum value $\bar{\gamma}$ of the update gain $\gamma$  in Eq.~\eqref{system_non_source} can be found from Lemma~\ref{lem_Stability_and_Update_gain} as $\bar{\gamma}=0.2763$. 
The update gain $\gamma$ needs to be smaller than the maximum value to ensure stability without DSR, and therefore,  the following simulations use the update gain
$$\gamma  = 0.1382  = \frac{\bar{\gamma}}{2} < \bar{\gamma}.$$
The discrete time system without DSR in Eq.~\eqref{system_non_source}   
settles to $2\%$ of the final value in $k_{Ts} = 1331$ steps, and for a settling time of 
$T_s = 1$ s, the sampling time is $ \delta_t = T_s/k_{Ts} = 7.5131 \times 10^{-4}$.

\subsection{Performance without and with DSR}
\begin{figure}[!ht]
\begin{center}
\includegraphics[width=.75\columnwidth]{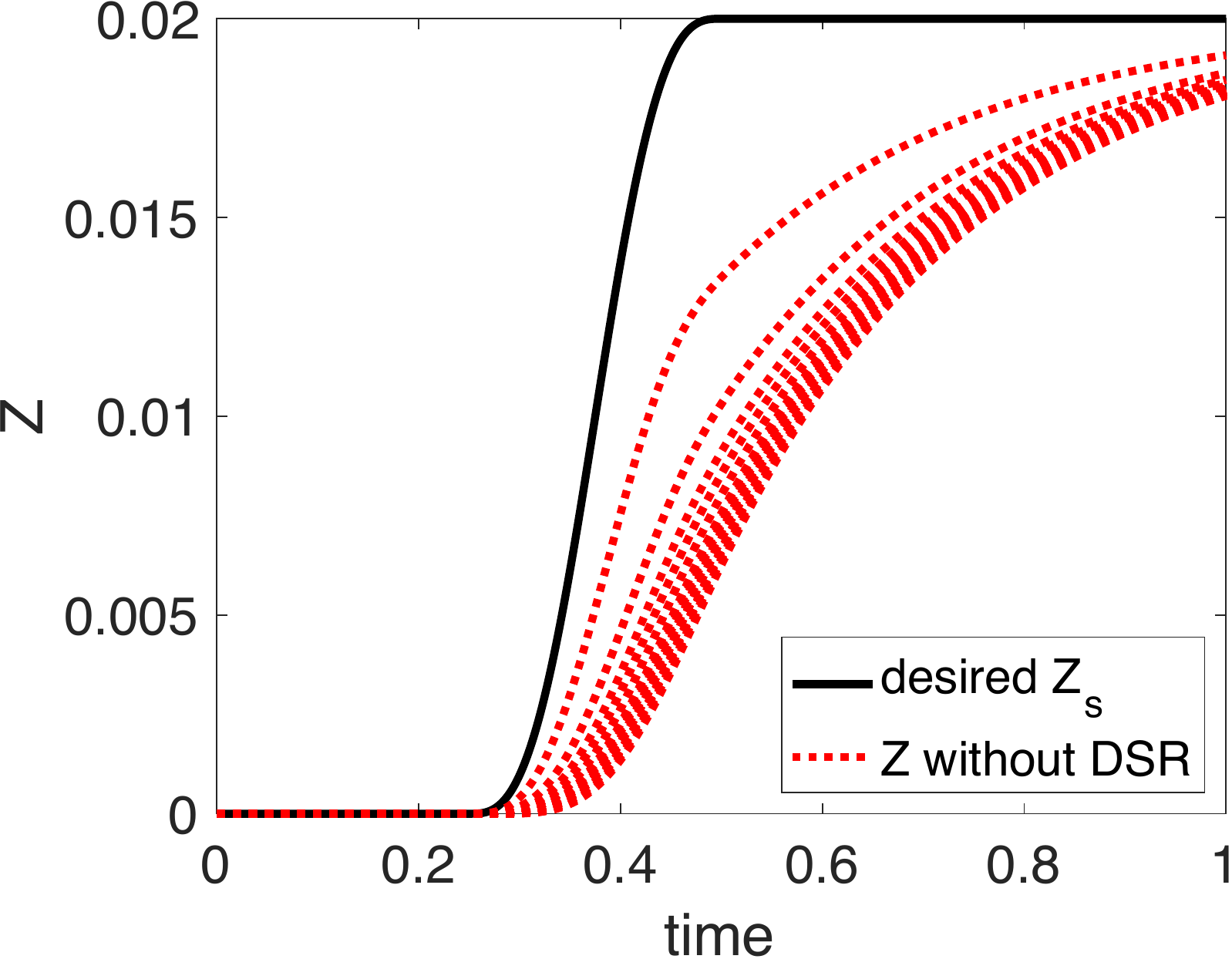}
\vspace{0.01in}
\caption{Response $Z$ of the standard consensus approach (without DSR) in Eq.~\eqref{system_non_source}, and the desired value  $Z_s$ at the source. 
}
\label{Fig_states_noDSR}
\end{center}
\end{figure}

The desired velocity of the source $Z_s$ is selected to increase, with a sinusoidal acceleration profile, from zero to  the desired consensus value of $$Z_d = 0.02$$ as shown in Fig.~\ref{Fig_states_noDSR}. 
The response of the states $Z$ achieves the final desired value $Z_d$ with a settling time of 
$T_s =1$s to reach and stay within $2\%$ of the final value $Z_d$, as shown in Fig.~\ref{Fig_states_noDSR}. 

\suppressfloats

\begin{figure}[!ht]
\begin{center}
\includegraphics[width=.85\columnwidth]{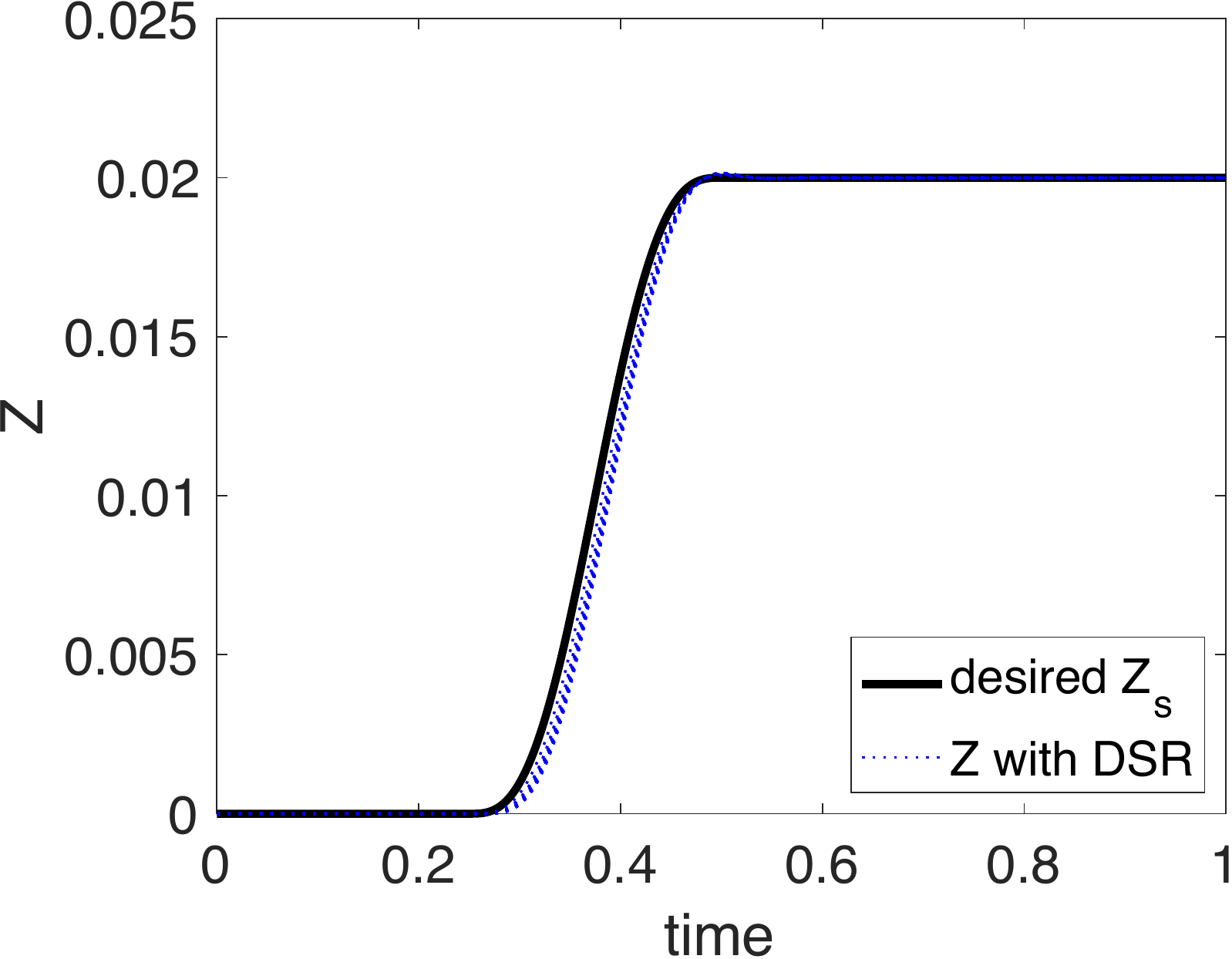}
\vspace{0.01in}
\caption{Response $Z$ of the accelerated-consensus approach implemented with DSR 
as in Eq.~\eqref{accelerated_system_non_source},  for the same desired source  time-profile  $Z_s(t)$ 
as in Fig.~\ref{Fig_states_noDSR} 
}
\label{Fig_states_DSR}
\end{center}
\end{figure}

\vspace{0.1in}
The response with DSR is substantially faster when compared to the response without 
the DSR, for the same desired source $Z_s$. 
With the  DSR gain selected as $\beta = 0.95$ in Eq.~\eqref{accelerated_system_non_source}, the settling time  is 
$T_s =0.4756$s, i.e.,  to reach and stay within $2\%$ of the final value $Z_d$, as shown in Fig.~\ref{Fig_states_DSR}. Thus, with a smaller settling time, the response with the DSR-based 
accelerated consensus is about $50\%$ faster than the response without 
DSR, as also seen by comparing the responses in Figs.~\ref{Fig_states_noDSR} and \ref{Fig_states_DSR}. 

\suppressfloats

\newpage
In addition to increasing the rate of convergence, and more importantly, 
the accelerated consensus leads to better synchronization during the transition. The 
deviation $\Delta$ (from synchronization) as in Eq.~\eqref{Eq_synchronization} of the response without the accelerated consensus as in Eq.~\eqref{system_non_source}
is $\Delta = 0.0103$. The normalized deviation $\Delta^*$ in Eq.~\eqref{Eq_norm_synchronization} 
is the same $\Delta^* =\Delta/T_s  = 0.0103$
since the settling time is $T_s=1$ without DSR.
The use of the accelerated consensus reduces the loss of synchronization during the
transition. The deviation   $\Delta= 0.0006$ for the accelerated consensus case. 
Even the normalized deviation (with a smaller settling time $T_s = 0.4756$) 
for the accelerated consensus using DSR is $\Delta^* =0.0012$, which is about ten times smaller than the case without the DSR. 

\vspace{0.1in}
Thus, the proposed accelerated-consensus approach with DSR can substantially improve synchronization during the transition by about ten times, in addition 
to decreasing the transition time by about half, 
when compared to the case without the DSR approach.

\subsection{Impact on formation spacing}
\begin{figure}[!ht]
\begin{center}
\includegraphics[width=.75\columnwidth]{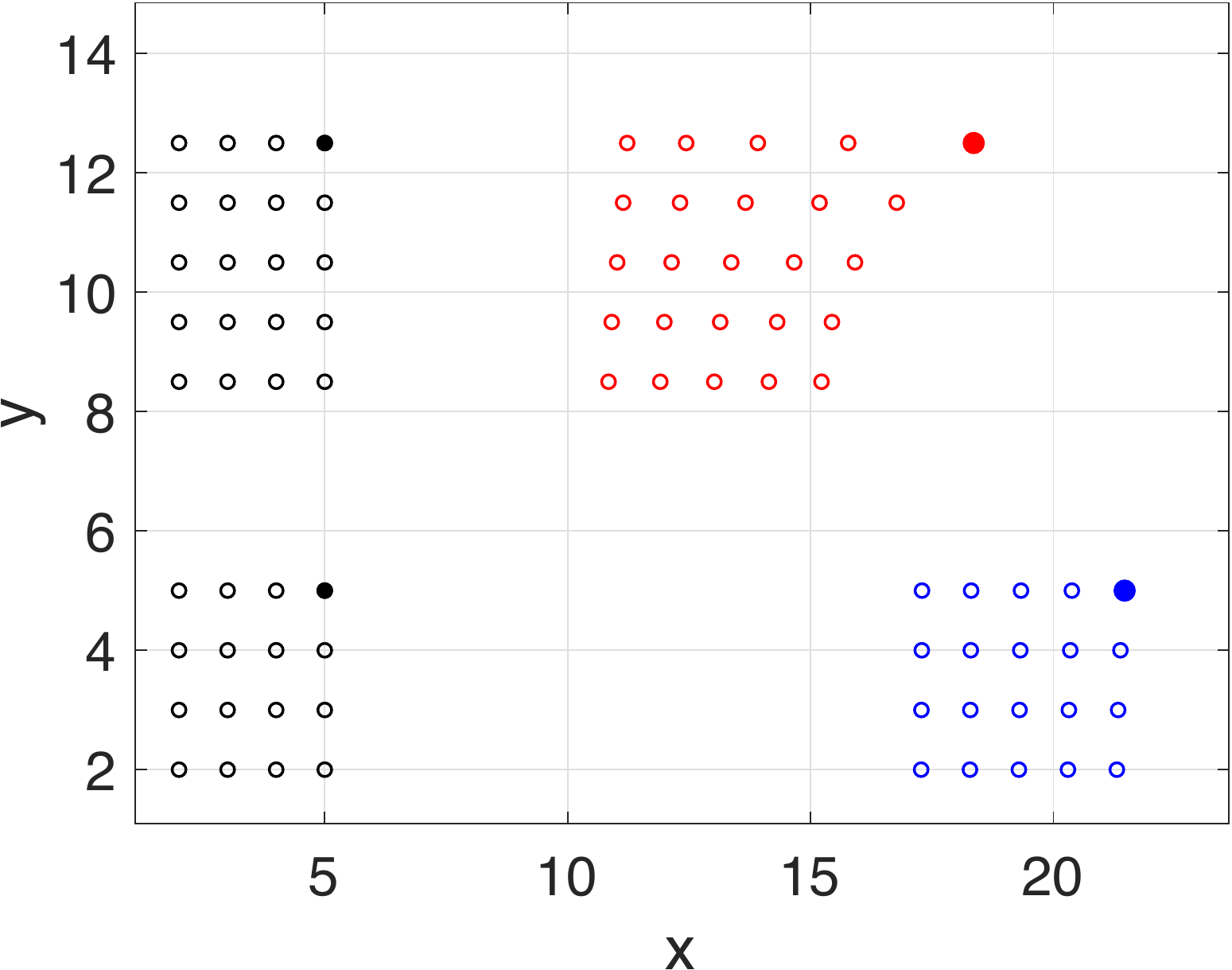}
\vspace{0.01in}
\caption{Accelerated consensus with DSR leads to better formation control 
when compared to the case without the DSR approach. Top (above $y=8$) shows the case without DSR with initial 
position in black and final position in red. Bottom (below $y=6$) shows 
the case with  DSR-based accelerated consensus,  with initial 
position in black and final position in blue.
}
\label{Fig_comp}
\end{center}
\end{figure}

To comparatively evaluate the impact of maintaining synchronization during the 
transition between consensus values,  the 
state $Z$ of the agents are considered to represent the horizontal velocity $Z=V_x$ of each agent. Then, the horizontal position $X$ of each agent is found as 
\begin{align}
X[k+1] &  = X[k] + \delta_t V_x[k]~  \nonumber  \\
& = X[k] + \delta_t Z[k].
\label{system_discrete}
\end{align}
The initial and final positions with and without the accelerated consensus are
compared in Fig.~\ref{Fig_comp}. 
As seen in the figure, the accelerated-consensus approach implemented with DSR as in Eq.~\eqref{accelerated_system_non_source} 
leads to better formation control 
when compared to the case without the DSR approach as in Eq.~\eqref{system_non_source}.

\suppressfloats

\vspace{0.1in}
\newpage
In this example, no control actions are taken to maintain the formation to focus on the comparative evaluation of the performance with and without the proposed  
accelerated-consensus approach. 
Nevertheless, the ability of the accelerated-consensus approach to reduce distortions in the formation can potentially improve the performance of other methods with control actions to maintain the formation.

\subsection{Summary of results} 
The use of the accelerated consensus, implemented using DSR,  results in a faster convergence to the consensus value. Moreover, during the transition the network is more cohesive with the accelerated-consensus approach, which results in better formation control. 
While this article focussed on a quadratic potential with a linear network dynamics, the accelerated-consensus approach could also be implemented for the nonlinear case.

\section{Conclusions}
\vspace{-0.01in}
This article showed that accelerated-gradient methods, used to improve the 
convergence in gradient-based search algorithms,  can be used to improve 
current consensus algorithms in networked multi-agent systems. 
Moreover, the article developed implementation of the proposed 
accelerated consensus  using delayed self reinforcement (DSR), 
where each agent only uses 
current and past information from the network. This is advantageous since the 
consensus improvement is achieved without the need to change the network connectivity 
and without the need for additional information from the network. 
Simulation results showed that the proposed accelerated-consensus approach with DSR can substantially improve synchronization during the transition by about ten times, in addition 
to decreasing the transition time by about half, 
when compared to the case without the DSR approach.
This was shown to improve formation control during transitions in networked 
multi-agent systems.


\end{document}